\documentclass[showpacs,prb,onecolumn]{revtex4}
\usepackage{bm}
\usepackage{graphicx}
\usepackage{amsmath}
\usepackage{amssymb}
\usepackage{eufrak}
\usepackage{color}
\newcommand{\nix}[1]{}

\begin{document}
\title{Amplification of spin-filtering effect by magnetic field in GaAsN alloys }

\author{V.~K.~Kalevich}
\author{M.~M.~Afanasiev}
\author{A.~Yu.~Shiryaev}
\author{A.~Yu.~Egorov} \affiliation{A.~F.~Ioffe Physico-Technical
Institute, St. Petersburg 194021, Russia}

\begin{abstract}
Abstract: We have found that intensity $I$ and circular
polarization degree $\rho$ of the edge photoluminescence, excited
in GaAsN alloys by circularly polarized light at room temperature,
grow substantially in the longitudinal magnetic field $B$ of the
order of 1\,kG. This increase depends on the intensity of pumping
and, in the region of weak or moderate intensities, may reach a
twofold value. In two-charge-state model, which considers
spin-dependent recombination of spin-oriented free electrons on
deep paramagnetic centers, we included the magnetic-field
suppression of spin relaxation of the electrons bound on centers.
The model describes qualitatively the rise of $\rho$ and $I$ in a
magnetic field under different pump intensities. Experimental
dependences $\rho(B)$ and $I(B)$ are shifted with respect to zero
of the magnetic field by a value of $\sim$170\,Gauss, while the
direction of the shift reverses with change of  the sign of
circular polarization of pumping. As a possible cause of the
discovered shift we consider the Overhauser field, arising due to
the hyperfine interaction of an electron bound on a center with
nuclei of the crystal lattice in the vicinity of the center.
\end{abstract}


\maketitle

\section{Introduction}
In the last few years the spin properties of GaAsN alloys have
aroused great interest\cite{Okayama2003, Egorov2005, JETPLett2005,
Lombez2005, JETPLett2007, Toulouse2007,obzor2008, SST2008,
physica2009, Nature2009, Zhao2009, APL2009, Toulouse2009,
China2009, JPCM2010, Toulouse2010, Linkop2011}, since, under
optical spin pumping, these alloys reveal an extremely high spin
polarization of free electrons (up to $90\%$) and its time
preservation ($\sim$1 ns) at room temperature\cite{Okayama2003,
Egorov2005, JETPLett2005, Lombez2005}. The findings look promising
for those compounds to be used in proposals for quantum
computation and spintronics. The anomalous enhancement of
free-electron polarization and its time conservation in GaAsN are
due to the dynamic polarization of electrons bound on deep
paramagnetic centers\cite{JETPLett2005,JETPLett2007,Toulouse2007}.
The paramagnetic centers arise with incorporation of nitrogen in
GaAs\cite{Nature2009,APL2009} and are polarized as a result of
spin-dependent capture of the polarized conduction electrons on
them. The polarized centers act as a spin filter, blocking the
capture of the spin-majority free electrons and promoting the
spin-minority photoelectrons to disappear from the conduction
band. The spin-filter efficiency rises with excitation light
intensity and can increase free electron polarization up to
$\approx100\%$ under strong pumping. Besides of the increase of
free-electron spin polarization, the accumulation of the
spin-majority free electrons in the conduction band leads to a
strong enhancement of the edge photoluminescence (PL) intensity
(up to 8 times) \cite{JETPLett2005,APL2009,JPCM2010}, as well as
the photoconductivity\cite{Toulouse2009,Toulouse2010}. At present,
the behavior of the nonlinear system of coupled spin-polarized
free and bound electrons under different conditions is adequately
investigated experimentally and analyzed theoretically, which
makes it possible to determine all the nine parameters of the
system\cite{JPCM2010}. However, these studies were all carried out
either in a zero- or in a perpendicular magnetic field (Voigt
geometry).

The present work investigates experimentally and theoretically the
influence of a longitudinal magnetic field (Faraday geometry) on
the coupled spin system of free and bound electrons under
different intensities of continuous-wave (cw) optical pumping. We
have found that the application of a longitudinal magnetic field
of the order of 1\,kG brings about a substantial (up to twice as
much) increase in the circular polarization and, at the same time,
in the edge PL intensity. The basis of this effect is the
suppression of spin relaxation of localized bound electrons by the
longitudinal field. In other words, the longitudinal magnetic
field increases the efficiency of the spin filter set up by
spin-polarized localized electrons. This amplification becomes the
greatest at low or moderate pumping intensities, when the
efficiency of spin-filter effect is rather small in the absence of
the magnetic field. Under great intensities of pumping when even
in a zero magnetic field the spin-dependent recombination leads to
a practically full spin polarization of free and bound electrons,
the influence of the longitudinal magnetic field weakens
dramatically. In order to describe the increase of $\rho$ and $I$
theoretically, we have introduced phenomenologically the
suppression of spin relaxation of bound electrons into the
two-charge-state model\cite{JETPLett2005, JETPLett2007, JPCM2010},
earlier used with success to simulate basic experiments on
spin-dependent recombination (SDR) in GaAsN in the zero- and the
perpendicular magnetic field. The calculation results describe
qualitatively the experimentally observed growth of $\rho$ and $I$
in a longitudinal magnetic field under different pump intensities.

We have also found that the experimental dependences of $\rho$ and
$I$ on the longitudinal magnetic field are shifted by a value of
$\sim170$\,Gauss with respect to zero of the field, the direction
of the shift changing to the opposite when the sign of circular
polarization of the exciting light changes. As a possible cause of
such shifts, we consider the hyperfine interaction of paramagnetic
centers with neighbouring nuclei of crystal lattice.

\section{Samples and experimental details} We studied the undoped $0.1\,\mu$m-thick
GaAsN layer with nitrogen content of 2.1\% grown by
rf-plasma-assisted molecular-beam epitaxy at 350--450$^{\circ}$C
on semi-insulating (001) GaAs substrate\,\cite{Egorov2005}. The
as-grown structure was annealed for 5\,min at 700$^{\circ}$C in a
flow of arsenic in the growth chamber. Free-electron spin
polarization  $P$ was created upon the interband absorption of
circularly polarized light \cite{OO,OO2}. We measured the
steady-state degree of circular polarization of the edge PL, which
is proportional to free-electron polarization \cite{OO, OO2}:
$\rho=P'P$, where the numerical factor $P'\leq1$, $\rho$ is
defined as $\rho=(I^+ -I^-)/I$, $I^+$ and $I^-$ are the right
($\sigma^+$) and left ($\sigma^-$) circularly polarized PL
components, $I=(I^+ +I^-)$ is the total PL intensity.
Continuous-wave tunable Ti:sapphire laser was used for
photoluminescence excitation. The PL was dispersed by a
monochromator and detected by a photomultiplier with an InGaAsP
photocathode. The $\rho$ and $I$ values were measured using a
high-sensitive polarization analyzer \cite{Kulkov} comprising a
quartz polarization modulator \cite{Jasperson} operating at
34\,kHz and a lock-in two-channel photon counter. The measurements
were carried out at 300\,K under normal incidence of the circular
polarized laser beam onto the sample and the detection of
luminescence in the opposite direction (backscattering
configuration).

\begin{figure}
  \centering
    \includegraphics[width=0.35\linewidth]{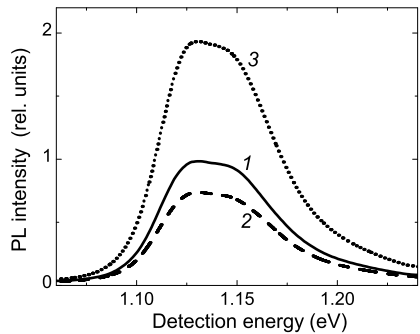}
    \caption{\label{I(SpB)} Room temperature PL spectra of GaAs$_{0.979}$N$_{0.021}$
    layer measured in zero magnetic field under circular (curve \emph{1}) and linear (curve
    \emph{2}) polarized pumping and in the longitudinal magnetic field of 3.5\,kG
    under circular polarized excitation (curve \emph{3}).
 Excitation energy $\hbar\omega_{\rm exc}=1.393$\,eV (excitation
below the GaAs barrier), excitation intensity $J=75$\,mW. }
\end{figure}

\section{Experimental results}
Figure\,\ref{I(SpB)} shows the spectra of PL intensity measured at
room temperature in GaAs$_{0.979}$N$_{0.021}$ layer in zero
magnetic field under linear (dashed curve) and circular (solid
curve) polarized excitation and also in a longitudinal magnetic
field of 3.5\,kG under circularly polarized pumping (dotted
curve). The PL intensity in zero field grows significantly (by
35\%) with polarization of pumping changed from the linear (dashed
curve) to the circular one (solid curve), pointing clearly to a
spin-dependent capture of conduction electrons onto deep
paramagnetic centers, which brings about the dynamic spin
polarization of centers and generation of spin-filtering effect
under circularly polarized excitation\cite{JETPLett2005,
Nature2009, APL2009}. The application of a relatively small
longitudinal magnetic field of 3.5\,kG under circularly polarized
pumping leads to an additional strong (about two times)
enhancement of PL intensity (dotted curve). This gives evidence of
the strengthening of spin-filtering effect by magnetic field. It
should be noted that the application of a longitudinal magnetic
field of the same value under linearly polarized pumping, when the
exciting light does not introduce the angular momentum in the
system of electrons, and the paramagnetic centers are left
non-polarized, does not change the PL intensity, so the measured
PL spectrum (not shown here) does not differ from curve \emph{2}
in Fig.\,\ref{I(SpB)}. The latter observation supports the
conclusion that the growth of $I$ in the magnetic field under
circular pumping is actually comes from the increase in the
efficiency of the spin filter caused by the longitudinal magnetic
field.

\begin{figure}
    \centering
    \includegraphics[width=0.3\linewidth]{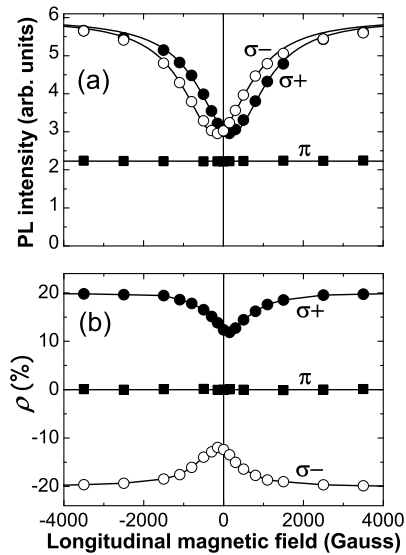}
    \caption{\label{IRo(B)} Dependences of the PL intensity (a) and PL circular polarization degree (b)
    on the longitudinal magnetic field measured at room temperature in GaAs$_{0.979}$N$_{0.021}$
    under excitation by right-circularly, $\sigma^{+}$ (solid circles), left-circularly, $\sigma^{-}$
    (open circles) and linearly, $\pi$ (solid squares) polarized light of the intensity
    $J=75$\,mW. Excitation energy is the same
    as in Fig.\,\ref{I(SpB)}, detection energy $\hbar\omega_{\rm
    det}=1.159$\,eV. }
\end{figure}

Figure\,\ref{IRo(B)}a shows the dependences of the PL intensity on
the longitudinal magnetic field measured near the PL band maximum
under excitation by right-circularly (solid circles),
left-circularly (open circles) and linearly (solid squares)
polarized light with intensity 75\,mW. These dependences differ
drastically for cases of the linearly- and the circularly
polarized pumping. Under linearly polarized excitation the PL
intensity is independent of the magnetic field. On the contrary,
under circularly polarized pumping the magnetic field brings about
a substantial growth of $I$, which becomes saturated in the field
$|B|\gtrsim2$\,kG, reaching a double value. Dependences $I(B)$ are
shifted with respect to zero of the field by a value
$|B_{\textrm{eff}}|\approx170$\,Gauss, and the sign of this shift
changes to the opposite with polarization of excitation changing
from $\sigma^{+}$ to $\sigma^{-}$. These dependences can be
approximated by Lorentz curves (solid curves in
Fig.\,\ref{IRo(B)}) of the form
$y(B)=y_{\textrm{max}}+(y_{\textrm{min}}-y_{\textrm{max}})/[1+(B-B_{\textrm{eff}})^2/B_{1/2}^2]$,
where $y_{\textrm{min}}=y(B=B_{\textrm{eff}})$,
$y_{\textrm{max}}=y(B\rightarrow\infty)$ and $B_{1/2}$ is the
half-width on the half-minimum of the curve. Later on we shall
return to discussion of physical origin of the effective field
$B_{\textrm{eff}}$, but now let us consider in detail the cause of
accretion of the PL intensity in a magnetic field. It is
known\cite{OO,OO2} that a longitudinal magnetic field suppresses
spin relaxation of electrons, leading to an increase of their
polarization. Indeed, the measured degree of circular polarization
of the edge luminescence $\rho$ (closed and open circles in
Fig.\,\ref{IRo(B)}b), which is directly proportional to the
polarization degree of free electrons $P$ under circular
($\sigma^{+}$ or $\sigma^{-}$) polarization of pumping, increases
in its absolute value (simultaneously with $I$ increasing) with
increase of the magnetic field. For comparison, note that under
linearly polarized excitation the value of  $\rho$ remains equal
to zero within the measurement error (closed squares in
Fig.\,\ref{IRo(B)}b). In the presence of spin-dependent
recombination the polarization of free electrons is strengthened
due to dynamic polarization of electrons bound on paramagnetic
centers. Therefore the observed growth of $|\rho|$ in the magnetic
field can arise as a result of suppression of the spin relaxation
of both free and bound electrons. However, the lifetime of free
electrons in the conduction band  $\tau$, which is determined by
rapid capture on deep centers and is in the order of magnitude of
1\,ps\,\cite{physica2009, JPCM2010}, is by two orders shorter than
the time of their spin relaxation
$\tau_{s}\sim150$\,ps\,\cite{Lombez2005,JETPLett2007,obzor2008},
and so an increase of $\tau_{s}$ in the longitudinal magnetic
field cannot involve the growth of $P$. On the contrary, the spin
relaxation time of bound electrons
$\tau_{sc}\sim1$\,ns\,\cite{physica2009} at the used pumping
intensities is comparable with their lifetime $\tau_{c}$, and even
shorter than that under weak pumping. Hence the increase of
$\tau_{sc}$ in a magnetic field must bring about an increase of
polarization of bound electrons $P_c$. Since the polarization
degree of bound electrons determines the efficiency of the spin
filter, the $P_c$ growth is accompanied by the increase, on one
hand, of free-electron polarization $P$ and, consequently, of
$\rho$, and on the other hand, of free-electron concentration in
the conduction band and, consequently, of $I$.

\begin{figure}
    \centering
    \includegraphics[width=0.35\linewidth]{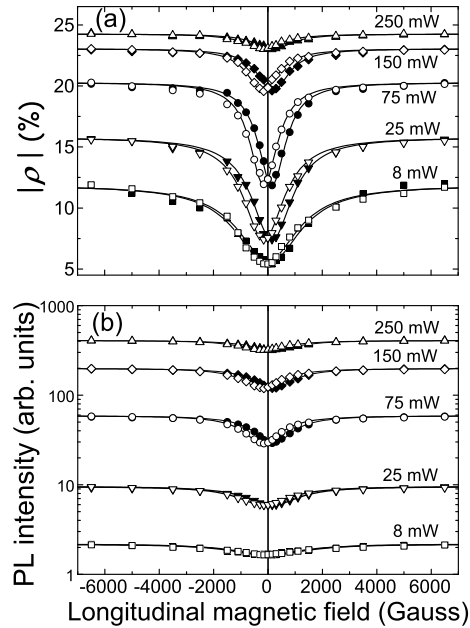}
    \caption{\label{IRo(BS+-)} PL circular polarization degree (a) and PL intensity
    (b) as functions of a longitudinal magnetic field measured in
GaAs$_{0.979}$N$_{0.021}$ for different intensities $J$ of the
right-circularly (solid symbols) and left-circularly (open
symbols) polarized pumping. $J$ (in mW):~ $\blacksquare, \square
-8;~\blacktriangledown, \triangledown-25;~ \bullet, \circ-75;~
\blacklozenge, \lozenge-150;~ \blacktriangle, \vartriangle-250$.
Solid curves are the result of fitting by Lorentzians
$y(B)=y_{\textrm{max}}+(y_{\textrm{min}}-y_{\textrm{max}})/[1+(B-B_{\textrm{eff}})^2/B_{1/2}^2]$,
where $y_{\textrm{min}}=y(B=B_{\textrm{eff}})$.~$T=300$\,K,
$\hbar\omega_{\rm exc}=1.393$\,eV, $\hbar\omega_{\rm
det}=1.159$\,eV. }
\end{figure}

This explanation finds an additional confirmation in consideration
of experimental dependences  $|\rho(B)|$ and $I(B)$, measured at
different intensities $J$ of the $\sigma^{+}$ (solid symbols) and
$\sigma^{-}$ (open symbols) exciting light and plotted in
Fig.\,\ref{IRo(BS+-)}. All these dependences grow with magnetic
field growing and reach saturation in the field $|B|\sim2$\,kG,
while having minima shifted with respect to zero magnetic field by
a value $|B_{\textrm{eff}}|\sim170$\,Gauss. Note that the
dependences for $\sigma^{-}$ pumping, being reflected in the plane
relative to the axis of ordinates, coincide within a few metering
errors with the curves recorded under $\sigma^{+}$ excitation.
Attention is attracted to the fact that the saturation level and
the extent of the increase of $\rho$, as well as $I$, depend on
pumping intensity nonlinearly. The dependence of the saturation
level of polarization $\rho(B_{\textrm{max}})$ (where
$B_{\textrm{max}}=6.5$\,kG is the maximal applied magnetic field)
on $\sigma^{+}$ excitation power is shown by closed circles in
Fig.\,\ref{RoI(J)}a, together with polarization dependence in zero
field $\rho(B=0)$ (open circles). It is seen that
$\rho(B_{\textrm{max}})$ is significantly larger than $\rho(B=0)$
for the intensities in the region of $10\div100$\,mW, while it
approximates $\rho(B=0)$ under the strong and weak pumpings.

The dependences of maximal increases of $\rho$ and $I$ induced by
magnetic field on pumping intensity can be easily followed,
considering normalized curves $\rho(B_{\textrm{max}})/\rho(B=0)$
(open squares) and $I(B_{\textrm{max}})/I(B=0)$ (solid squares) in
Fig.\,\ref{RoI(J)}b. We see that each of these curves has a
maximum (at $J\approx15$\,mW and 70\,mW respectively) and falls
drastically on both sides from it, tending to unity within the
limit of strong and weak pumpings. Such non-monotonous influence
of the magnetic field on intensity and polarization of PL fits in
naturally with the picture of spin-dependent recombination.
Indeed, the decrease of $J$ down to vanishingly small values leads
to disappearance of the spin-filter effect\cite{JPCM2010}, which
must decrease the relationships $I(B_{\textrm{max}})/I(B=0)$ and
$\rho(B_{\textrm{max}})/\rho(B=0)$ to unity. At large $J$ (in our
case at $J>>70$\,mW) the values of $I(B_{\textrm{max}})/I(B=0)$
and $\rho(B_{\textrm{max}})/\rho(B=0)$ must also tend to unity,
but for the opposite reason: under strong pumping the spin-filter
effect provides a practically complete polarization of centers
(and, simultaneously, free electrons) even in zero magnetic
field\cite{JPCM2010}, therefore the application of a longitudinal
magnetic field cannot increase $P_c$ (and $P$) and, as a
consequence, the polarization and intensity of PL.

\section{Model and comparison with experiment}
For theoretical description of changes in $I$ and $\rho$ in a
longitudinal magnetic field we have supposed that this field
suppresses the spin relaxation of the electrons localized on
centers. According to Ref.\,[\cite{DP1973}] (see also Chapter 2 in
Ref.\,[\cite{OO}] and Chapter 1 in Ref.\,[\cite{OO2}]), spin
relaxation of localized electrons can be presented as a result of
the action of chaotic local magnetic fields on their spins. The
application of an external magnetic field $B$ parallel to the
exciting beam (the mean spin of photoexcited electrons) suppresses
the action of those local fields. This suppression becomes
significant when the magnitude of magnetic field equals the
characteristic value $B_f$ of the chaotic field or exceeds it. The
increase of the spin relaxation time of localized electrons
$\tau_{sc}$ in an external magnetic field $B$ is described by the
expression\cite{DP1973}:
\begin{equation} \label{tau}
\frac{1}{\tau_{sc}(B)} =
   \frac{ \frac{2}{3} \omega_f^2 \tau_{cor}}{1+\omega^2
  \tau_{cor}^2} = \frac{1}{\tau_{sc}(0)} \frac{1}{1+(B/B_{1/2})^2} \: ,
\end{equation}
where $1/\tau_{sc}(0) =\frac{2}{3} \omega_f^2 \tau_{cor}$ is the
spin relaxation rate at $B=0$, $\omega_f=g_c\mu_B B_f/\hbar$  and
$\omega=g_c\mu_B B/\hbar$ are the Larmor frequencies of the
bound-electron spin in the random local magnetic field $B_f$ and
external magnetic field $B$, respectively, the correlation time
$\tau_{cor}$ is the characteristic time of changing the field
$B_f$, $g_c$ is the bound-electron \emph{g}-factor, $\mu_B$ is the
Bohr magneton, and $B_{1/2}=\hbar/(g_c\mu_B \tau_{cor})$ is the
magnitude of the external field in which the spin relaxation time
doubles. Equation\,(\ref{tau}) is obtained on the assumption of
short correlation time when $\omega_f \tau_{cor} <<1$. According
to Eq.\,(\ref{tau}), in a strong field $B>>B_{1/2}$ the spin
relaxation is suppressed totally, independent of pumping
intensity. Consequently, polarization of electrons must attain its
maximum value, the same for any intensity of excitation. At the
same time, as is seen from Fig.\,\ref{IRo(BS+-)}a, the saturation
level of PL polarization in a large magnetic field rises with
increasing pumping. This is possible if there is an additional
channel of spin relaxation, which is not suppressed by the used
magnetic field $|B|\leq6.5$\,kG. The additional relaxation can be
taken into account by introducing the term $1/\tau_{sc}^*$ into
Eq.\,(\ref{tau}), where $\tau_{sc}^*$ is the spin relaxation time
independent of the magnetic field:
\begin{equation} \label{tau1}
\frac{1}{\tau_{sc}(B)} = \frac{1}{\tau_{sc}^*}+
\frac{1}{\tau_{sc}^{(1)}} \frac{1}{1+(B/B_{1/2})^2} \: ,
\end{equation}
$1/\tau_{sc}^{(1)} = [1/\tau_{sc}(0)]- 1/\tau_{sc}^*$.

\begin{figure}
  \centering
    \includegraphics[width=0.3\linewidth]{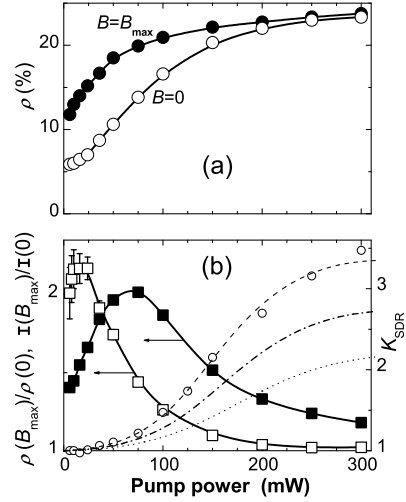}
    \caption{\label{RoI(J)} Experimental dependences $\rho(B=0)$,
    $\rho(B=B_{\textrm{max}})$ (a) and $\rho(B_{\textrm{max}})/\rho(0)$ (open squares),
    $I(B_{\textrm{max}})/I(0)$ (solid squares) and $K_{\textrm{SDR}}$ (open circles) (b) on
    the right-circular pump power, $J$, where
    $B_{\textrm{max}}=6.5$\,kG. Solid curves are the guides for the eye. Dashed, dotted and
    dashed-dotted lines are the result
    of model calculation (see text for details). }
\end{figure}

As it has been demonstrated in Ref.\,[\cite{JPCM2010}], the
two-charge-state model of SDR, based on assumption that the deep
center may be occupied by one (paramagnetic state) or two (singlet
state with zero spin) electrons\,\cite{Lampel}, describes well the
basic experimental results on spin dynamics in GaAsN in zero or a
perpendicular magnetic field. In order to take into account the
effect of the longitudinal magnetic field we have used the same
model, introducing into the rate equations (4) of
Ref.\,[\cite{JPCM2010}] the increase of spin relaxation time of
localized electrons, given by Eq.\,(\ref{tau1}). Besides
$\tau^*_{sc}$ and $B_{1/2}$, the model contains 9 parameters, and
for eight of them we have retained the values found in
Ref.\,[\cite{JPCM2010}] for the same sample, viz.,
$\tau_{sc}(0)=700$\,ps, the spin relaxation time of free electrons
$\tau_s=140$\,ps, the minimum lifetime of free electrons
$\tau^*=2$\,ps, the minimum lifetime of free holes
$\tau^*_h=30$\,ps, Lander \emph{g}-factors of free and bound
electrons \emph{g}=+1 and $g_c=+2$, respectively, the density of
deep centers $N_c=3\times10^{15}\,\textrm{cm}^{-3}$, and
$P'=0.28$, where $P'$ is  the factor relating $\rho$ with $P$
($\rho=P'P$). The ninth parameter, the initial polarization of
electrons $P_i$, depends on the energy of exciting light quanta
$\hbar\omega_{\rm exc}$. We cannot use the value $P_i=0.24$ from
Ref.\,[\cite{JPCM2010}], since it was found at $\hbar\omega_{\rm
exc}=1.312$\,eV, while in the present work we have used much
greater energy $\hbar\omega_{\rm exc}=1.393$\,eV. In
GaAs$_{0.979}$N$_{0.021}$ the excitation energy $\hbar\omega_{\rm
exc}=1.393$\,eV is close to the energy of optical transition from
the spin-orbit split-off valence band to the conduction band. For
this transition the spin polarization of the electrons \emph{in
statu nascendi} has the opposite sign, which can lead to a
significant decrease of $P_i$\cite{OO,OO2}. A strong decrease of
$P_i$ is observed in GaAsN layers with [N]=1.3\% and 2.6\% as
$\hbar\omega_{\rm exc}$ increases up to values close to the band
gap of the GaAs barrier \cite{Linkop2011}.

\begin{figure}
  \centering
    \includegraphics[width=0.3\linewidth]{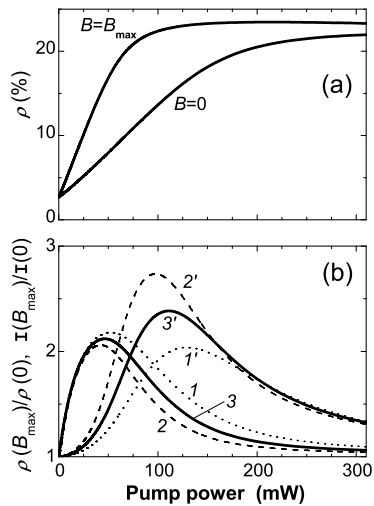}
    \caption{\label{calcRatio} Calculated dependences $\rho(B=0)$,
    $\rho(B=B_{\textrm{max}})$ (a) and $\rho(B_{\textrm{max}})/\rho(B=0)$ (curves \emph{1}, \emph{2} and
    $\emph{3}$), $I(B_{\textrm{max}})/I(B=0)$ (curves $\emph{1}'$, $\emph{2}'$ and $\emph{3}'$) (b)
    versus the right-circular polarized pump power for $P_i=0.08$ (dotted curves), $P_i=0.1$ (solid
    curves), $P_i=0.12$ (dashed curves)
    and $\tau_{sc}(B_{\textrm{max}})=\tau_{sc}^*=2200$\,ps (see text for details). }
\end{figure}

Our calculations have shown that the ratio between the height of
the maximum of the $I(B_{\textrm{max}},J)/I(0,J)$ curve and the
height of the maximum of the $\rho(B_{\textrm{max}},J)/\rho(0,J)$
curve depends heavily on the magnitude of $P_i$. For instance, it
decreases from $\approx3$ to $\approx1$ with $P_i$ decreasing from
0.24 to 0.08. The heights of maxima for the experimental curves
$I(B_{\textrm{max}},J)/I(0,J)$ and
$\rho(B_{\textrm{max}},J)/\rho(0,J)$ in Fig.\,\ref{RoI(J)}b are
about equal, which allows us to suppose that $P_i$ is close to
0.08. It is evident that an increase in time $\tau^*_{sc}$
increases the steepness of the build-up of the
$\rho(B_{\textrm{max}})$ (and also $I(B_{\textrm{max}})$)
dependence  and, respectively, the height of the maximum of the
$\rho(B_{\textrm{max}})/\rho(0)$ (and also
$I(B_{\textrm{max}})/I(0)$) dependence on $J$. Analysis has shown
that the influence of $\tau^*_{sc}$ on the height of maxima is
rather weak. The heights of maxima for the calculated curves turn
out equal to the heights of maxima of the respective experimental
dependences in Fig.\,\ref{RoI(J)}b at $P_i=0.08$ and
$\tau_{sc}(B_{\textrm{max}})=\tau_{sc}^*=2200$\,ps (see dotted
curves \emph{1} and \emph{1}$'$ in Fig.\,\ref{calcRatio}b.

As shown in Ref.\,[\cite{JPCM2010}], the $P_i$ value can be
determined independently from a model fitting of an experimental
dependence of the SDR ratio $K_{\textrm{SDR}}=I(\sigma,
B=0)/I(\pi, B=0)$ on the pump power, where $I(\sigma, B=0)$ and
$I(\pi, B=0)$ is the PL intensity in zero magnetic field under
circularly and linearly polarized excitation, respectively. The
experimental curve $K_{\textrm{SDR}}(J)$ recorded at
$\hbar\omega_{\rm exc}=1.393$\,eV is shown in Fig.\,\ref{RoI(J)}b
by open circles\cite{rem1}. Also presented here are the
dependences $K_{\textrm{SDR}}(J)$, calculated for $P_i= 0.08$,
0.1, and 0.12 (dotted, solid, and dashed curves, respectively). It
is seen that the curve calculated for $P_i= 0.08$ (dotted) runs
appreciably below the experimental curve. The coincidence with
experimental curve occurs for $P_i= 0.12$ (dashed). In this case,
however, the ratio of maxima for the calculated curves
$I(B_{\textrm{max}},J)/I(0,J)$ and
$\rho(B_{\textrm{max}},J)/\rho(0,J)$ (dashed curves \emph{2} and
\emph{2}$'$ in Fig.\,\ref{calcRatio}b) is markedly more than 1.
The optimal agreement between the calculation results and the
experimental data in Fig.\,\ref{RoI(J)} is obtained for $P_i= 0.1$
(see dashed-dotted curve in Fig.\,\ref{RoI(J)}b and solid curves
$\emph{3}$ and \emph{3}$'$ in Fig.\,\ref{calcRatio}).

\begin{figure}
  \centering
    \includegraphics[width=0.3\linewidth]{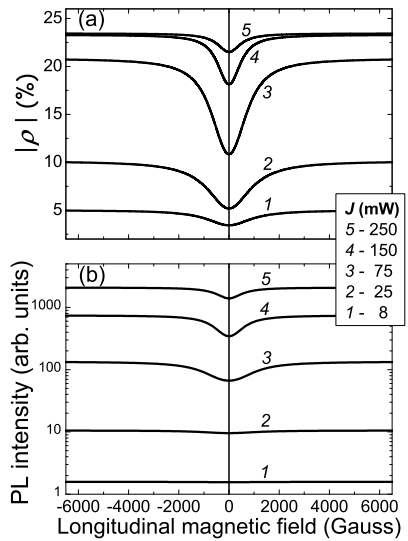}
    \caption{\label{calcIRo} Calculated dependences of $\rho(B)$
    (a) and $I(B)$ (b) for different $\sigma$ pump powers at
    $P_i=0.1$, $\tau_{sc}^*=2200$\,ps and $B_{1/2}=600$\,Gauss (see text for
    details). $J$ (in mW):~ $\emph{1}-8;~\emph{2}-25;~ \emph{3}-75;~
\emph{4}-150;~\emph{ 5}-250$. }
\end{figure}

The above determined values of the parameters $P_i= 0.1$ and
$\tau_{sc}^*=2200$\,ps were used for computer simulation of the
dependence of $\rho$ and $I$ on magnetic field. We evaluated the
value of the last, eleventh model parameter, necessary for that
computation, $B_{1/2}=600$\,Gauss, from fitting of the
experimental curves in Fig.\,\ref{IRo(BS+-)}a by Lorentzians.
Curves $\rho(B)$ and $I(B)$, calculated at different pump
intensities, are shown in Fig.\,\ref{calcIRo}. As is seen, they
describe qualitatively basic peculiarities of the experimental
curves in Fig.\,\ref{IRo(BS+-)}: 1) both $\rho$ and $I$ increase
in a magnetic field and reach saturation in a strong field
$|B|\gtrsim2$\,kG; 2) the saturation level of $\rho$ in a strong
field depends on excitation intensity, growing with growth in $J$;
3) the maximum increase of $\rho$ and $I(B)$ in magnetic field
depends nonmonotonically on pump intensity, decreasing under weak
or strong pumping. The latter is seen distinctly in
Fig.\,\ref{calcRatio}b, where calculated dependences
$\rho(B_{\textrm{max}},J)/\rho(0,J)$ (curve \emph{3}) and
$I(B_{\textrm{max}},J)/I(0,J)$ (curve \emph{3}$'$) have maxima and
tend to 1 at $J\rightarrow0$ and $J\rightarrow\infty$, so
confirming our qualitative argumentation.

Experimental dependences $\rho(B)$ and $I(B)$ in
Fig.\,\ref{IRo(B)} and Fig.\,\ref{IRo(BS+-)} are shifted
respective to zero of the field by a value
$|B_{\textrm{eff}}|\sim170$\,Gauss. The investigated GaAsN alloy
is a non-magnetic semiconductor, so the presence of such shifts
indicates that the circularly polarized pumping generates an
effective magnetic field, acting upon electron spins. This is
confirmed by the fact that the field $B_{\textrm{eff}}$ reverses
its direction when the sign of circular polarization of excitation
changes. The effective magnetic field, dependent on the sign of
polarization of pumping, is typical for low-temperature
experiments on optical orientation in semiconductors and
semiconductor nanostructures. This field is identified with the
field of dynamically polarized nuclei of crystal lattice
(Overhauser field), which arouses due to hyperfine interaction
with localized electrons (localized on shallow donors in a bulk
semiconductor, on fluctuations of the hetero-boundary potential in
a quantum well or in a quantum dot)\cite{OO,OO2}. Usually the rise
of crystal temperature delocalizes electrons, which destroys the
hyperfine interaction and, consequently, the Overhauser field. The
paramagnetic centers in GaAsN are deep and can maintain a strong
hyperfine interaction with lattice nuclei up to the room
temperature. Therefore we think that in our experiments the field
$B_{\textrm{eff}}$ is an Overhauser field of the crystal lattice
nuclei in the neighborhood of a paramagnetic center. Corroboration
of this conclusion and elucidation of the peculiarities of the
dynamic polarization of nuclei in the case of spin-dependent
recombination calls for additional experiments, which are in
progress.

It should be noted that the calculated curves $\rho(B)$ and $I(B)$
in Fig.\,\ref{calcIRo} are symmetric with respect to the ordinate
axis, since our model takes no account of the effective field
$B_{\textrm{eff}}$. Formally, the field $B_{\textrm{eff}}$ may be
taken into account in calculation through substitution of the
field $B$ by the total field $(B+B_{\textrm{eff}})$ where the sign
of $B_{\textrm{eff}}$ depends on the sign of circular polarization
of the exciting light.

\section{Conclusion}
In summary, under optical pump conditions the longitudinal
magnetic field of the order of 1\,kG increases the spin
polarization of electrons and the luminescence intensity in GaAsN
alloys at room temperatures. The increase in polarization, as well
as in PL intensity depends on the pump intensity an can reach
twice as much value. This effect can be used in problems of
spintronics to obtain optimal conditions for the maximum
polarization of free electrons, in particular, for intensification
of spin-dependent
photoconductivity\cite{Toulouse2009,Toulouse2010}. The modified
two-charge-state-model, which includes suppression of spin
relaxation of localized electrons by a longitudinal magnetic field
describes adequately the behavior of the coupled spin system of
free and bound electrons in GaAsN alloys in a longitudinal
magnetic field under different pump intensities.

We suppose that the chaotic magnetic fields, bringing about the
spin relaxation of GaAsN deep centers in zero magnetic field, are
the field fluctuations of nuclei\cite{Merc2002}, coupled with
paramagnetic centers by hyperfine interaction. Investigations
aimed at elucidation of the nature of chaotic fields are in
progress.

One of the most informative methods of studying centers and
defects in semiconductors is the optically detected electron
paramagnetic resonance (ODEPR). Its application, however, is
limited to low temperatures, where the spin lifetime for localized
electrons is long enough. For this reason, up to now the ODEPR
spectra of paramagnetic centers, responsible for SDR in GaAsN
alloys, have been measured only at liquid He
temperature\cite{Nature2009,APL2009}. The suppression of spin
relaxation of the centers in a longitudinal magnetic field with
simultaneous increase of their polarization up to a complete value
makes it possible to approach the conditions indispensable for
detection of the ODEPR of paramagnetic centers in GaAsN at higher
temperatures, room temperature including.

The effective magnetic field $B_{\textrm{eff}}$, that acts upon
localized electrons and changes the direction with the sign of
circular polarization of the exciting light changed, has been
detected at room temperature. We interpret this field as the
Overhauser field of the lattice nuclei optically oriented in the
vicinity of a paramagnetic center. Polarization of nuclei offers
new possibilities of controlling the spin state of free and bond
electrons at room temperature, for example, through its change
under conditions of nuclear magnetic resonance.

\acknowledgments{This research was supported by the programmes of
the Russian Academy of Sciences and the Russian Foundation for
Basic Research. We are grateful to E.L.~Ivchenko, K.V.~Kavokin and
L.S.~Vlasenko for fruitful discussions.}

\end{document}